\documentclass[%
 aip,
 rsi,
 amsmath,amssymb,
 reprint,%
]{revtex4-1}

\usepackage{graphicx}
\usepackage{dcolumn}
\usepackage{bm}

\usepackage[utf8]{inputenc}
\usepackage[T1]{fontenc}
\usepackage{mathptmx}

\begin{document}

\preprint{AIP/123-QED}

\title[Two-Color Differential Dynamic Microscopy]{Two-color differential dynamic microscopy for capturing fast dynamics}

\author{R. You}
\author{R. McGorty}%
 \email{rmcgorty@sandiego.edu.}
\affiliation{ 
Department of Physics and Biophysics, University of San Diego, San Diego, CA 92110, USA
}%

\date{\today}

\begin{abstract}
Differential dynamic microscopy (DDM) is increasingly used in the fields of soft matter physics and biophysics to extract the dynamics of microscopic objects across a range of wavevectors using optical microscopy. Standard DDM is limited to detecting dynamics no faster than the camera frame rate. We report on an extension to DDM where we sequentially illuminate the sample with spectrally-distinct light and image with a color camera. By pulsing blue and then red light separated by a lag time much smaller than the camera’s exposure time we are able to use this two-color DDM method to measure dynamics occurring much faster than the camera frame rate. 
The following article has been accepted by Review of Scientific Instruments. After it is published, it will be found at \url{https://aip.scitation.org/journal/rsi}.
\end{abstract}

\maketitle

A number of optical techniques allow users to quantify the dynamics of small particles, molecules, intracellular bodies or whole cells. These include single-particle tracking\cite{Crocker}, image correlation spectroscopy\cite{Petersen}, fluorescence correlation spectroscopy\cite{Magde}, and dynamic light scattering\cite{Berne}. Researchers were given another option to add to this list in 2008: differential dynamic microscopy (DDM)\cite{Cerbino}. Its advantages include being able to extract particle dynamics from time series of images even when the contrast is too low or the particle concentration is too high to allow for accurate particle localizations and being compatible with a number of optical microscopy modalities including bright-field, dark-field\cite{Bayles2016}, wide-field fluorescence, confocal\cite{Lu}, and light-sheet microscopies\cite{Wulstein}.

As DDM can be used with nearly any optical microscope with a digital camera and yields data analogous to what one would obtain with dynamic light scattering, it has been applied to numerous system. To provide just a partial list, DDM has been used to measure the dynamics of: colloidal particles or nanoparticles\cite{He}; anisotropic particles\cite{Reufer}; swimming bacteria\cite{Wilson}; biomacromolecules or particles in biomimetic environments\cite{Regan, Burla}; colloidal gels\cite{Cho}; probe particles for microrheological determinations of viscoelasticity\cite{Bayles2017}; particles in crowded environments\cite{Sentjabrskaja}; and foams\cite{Giavazzi}.

In this paper, we present a modification to DDM that allows dynamics faster than the camera frame rate to be measured. With standard DDM, one is unable to quantify dynamics occurring over times shorter than the time interval between camera frames. This makes studying fast dynamics problematic without access to high-speed cameras. We show that by using a color camera and illuminating the sample with spectrally-separated pulses of light one can uncover dynamics occurring faster than the camera frame rate. Inspiration for this method came from two-color particle velocimetry which has been used to measure fast dynamics\cite{Adrian, Goss}.

To use standard DDM, one acquires a time series of images taken using any spatially invariant microscopy method and analyzes the difference between images separated by a given lag time in Fourier space. One can then find the time it takes for intensity fluctuations to decay as a function of the spatial frequency or wavevector, $\boldsymbol{q}$. In practice, one first calculates the difference between images separated by a lag time $\Delta t$, $d(\boldsymbol{x},t,\Delta t)=I(\boldsymbol{x},t+\Delta t)-I(\boldsymbol{x},t)$ where $\boldsymbol{x}$ is the pixel position. One next obtains the image structure function by taking two dimensional Fourier transforms of these differences and averaging over all times $t$: $D(\boldsymbol{q},\Delta t)=\langle | \hat{d}( \boldsymbol{x},t,\Delta t)|^2 \rangle _t$. Isotropic samples will result in a radially symmetric image structure function which can then be azimuthally averaged to yield $D( q,\Delta t)$ where $q= \sqrt{q_x^2+q_y^2}$. This function can then be fit to
\begin{eqnarray}
D(q,\Delta t) = A(q) \big (1-f(q,\Delta t) \big ) + B(q),
\label{eq:one}
\end{eqnarray}
where the amplitude, $A(q)$, depends on the scattering properties of the sample and the optical properties of the microscope, the background, $B(q)$, depends on the noise in the image, and $f(q,\Delta t)$ is the intermediate scattering function which accounts for the sample dynamics. For diffusive dynamics $f(q,\Delta t)=\exp(-\Delta t/ \tau (q))$ where $\tau=(Dq^2)^{-1}$ and $D$ is the diffusion coefficient. 

Given that DDM analyzes images acquired with a digital camera on an optical microscope, the range of accessible spatial scales span from the diffraction limit or pixel size (whichever is greater) to the size of the field of view. In typical studies of thermally-diffusing colloids or moving bacteria, this range is often from the submicron to ~100 $\mu$m. Over what time scales will dynamics over this spatial range occur? For diffusive dynamics, the time for a density fluctuation to decay is proportional to the length scale squared. That is $\tau \propto q^{-2}$, where $\tau$ is the decay time and $q$ is the wavevector. Therefore, to probe diffusive dynamics across two orders of magnitude in space requires data over four orders of magnitude in time. 

How can one acquire data over such a span of timescales? In several DDM studies, image sequences are acquired at multiple frame rates. As one example, Germain et al. investigate the dynamics of micron-sized colloids and bacteria and acquire images at both 400 Hz and 4 Hz\cite{Germain}. This allows the authors to cover time scales from 2.5 ms to 1000 s. More recently, Arko and Petelin developed a dual-camera method to acquire DDM data over about 6 orders of magnitude in time\cite{Arko}. Using a beam splitter they imaged their sample onto two separate cameras, each recording frames at 200 Hz. By triggering the cameras at offset times and comparing frames between the two cameras, they could measure time lags much smaller than with a single camera. 

In this paper, we describe a new method which likewise provides access to dynamics faster than the frame rate of our camera. Using a single color camera and illuminating the sample with pulses of blue and red light, we can compare the blue and red channels of a single image to observe changes within the sample over times much smaller than the exposure time of a single frame. 

\begin{figure}[t]
\includegraphics[width=1\columnwidth]{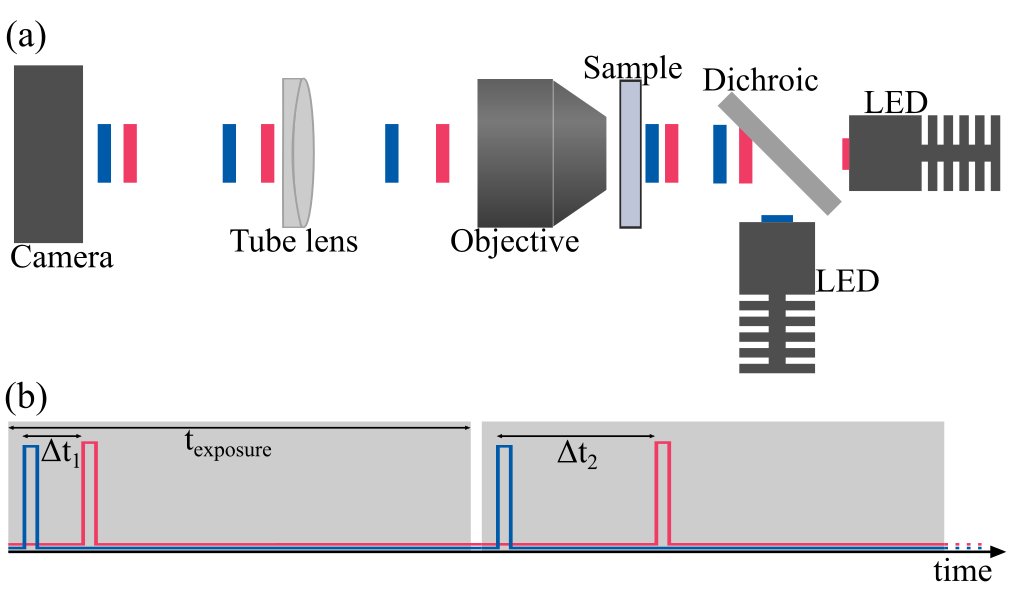}
\caption{\label{fig:fig1} (a) Schematic of our optical microscope shows two LED light sources (one with a peak wavelength of 625 nm, the other 455 nm) used to illuminate a sample. A 40× objective and tube lens image the sample onto a color CMOS camera. (b) Within a single exposure time of the camera, we pulse each LED. The time interval between pulses is varied and, for the data shown, ranges from 3 to 91 ms.}
\end{figure}

We perform two-color DDM measurements with the setup shown in Figure \ref{fig:fig1}a. The light from red and blue LEDs (M625L2 and M455L3, Thorlabs) are combined with a longpass dichroic mirror (DMLP550R, Thorlabs) and illuminate the sample. A 40× objective (0.65 NA, Olympus) and $f$ = 200 mm tube lens (AC254-200-ML-A, Thorlabs) image the sample onto a color CMOS camera (DFK 37BUX287, The Imaging Source). Triggering the two LEDs and the camera is a Digilent Analog Discovery 2 (National Instruments). 

The trigger sequencing is indicated in Figure \ref{fig:fig1}b. We trigger the camera to start an exposure every 100 ms, resulting in a frame rate of 10 Hz. Within a single exposure time, both the blue (455 nm) and red (625 nm) LEDs each turn on for 1 ms. The spacing between the blue and red pulses is staggered. We use a sequence of 10 pulse delays in our experiments: 3, 5, 7, 10, 14, 20, 29, 43, 62 and 91 ms. We repeat this sequence of different pulse delays 800 times. Therefore, we acquire 8000 color images acquired at 10 Hz. However, the short delay times between the blue and red illumination pulses allow us to measure dynamics occurring much faster than 10 Hz.

The analysis of this data follows closely to what was previously described for standard DDM. However, we now first separate the recorded time series of images into blue and red channels. We then calculate the image structure function using the differences between data in the blue and red channels: $d_{two-color} (\boldsymbol{x},t,\Delta t)=I_{red} (\boldsymbol{x},t+\Delta t)-I_{blue} (\boldsymbol{x},t)$. Here, $\Delta t$ no longer must be some integer multiple of the time between adjacent frames (as in standard DDM) but can be as small as the time separating the blue and red illumination pulses. 

We evaluate this method using a sample of 180-nm-diameter polystyrene particles (FluoSpheres F8811, Molecular Probes) suspended in water. These particles are diluted by a factor of 100 and a few microliters are sealed between a glass slide and coverslip. This sample was first investigated with standard DDM using a camera running at 80 Hz and constant bright-field illumination. We found a diffusion coefficient of $2.48 \pm 0.03 \mathrm{\mu m^2/s}$. We then acquired data using the two-color DDM method as described where we acquired 8000 frames of 256$\times$256 pixels with the camera running at 10 Hz and the blue and red illumination being pulsed with the times between pulses ranging from 3 to 91 ms.

\begin{figure}[t]
\includegraphics[width=1\columnwidth]{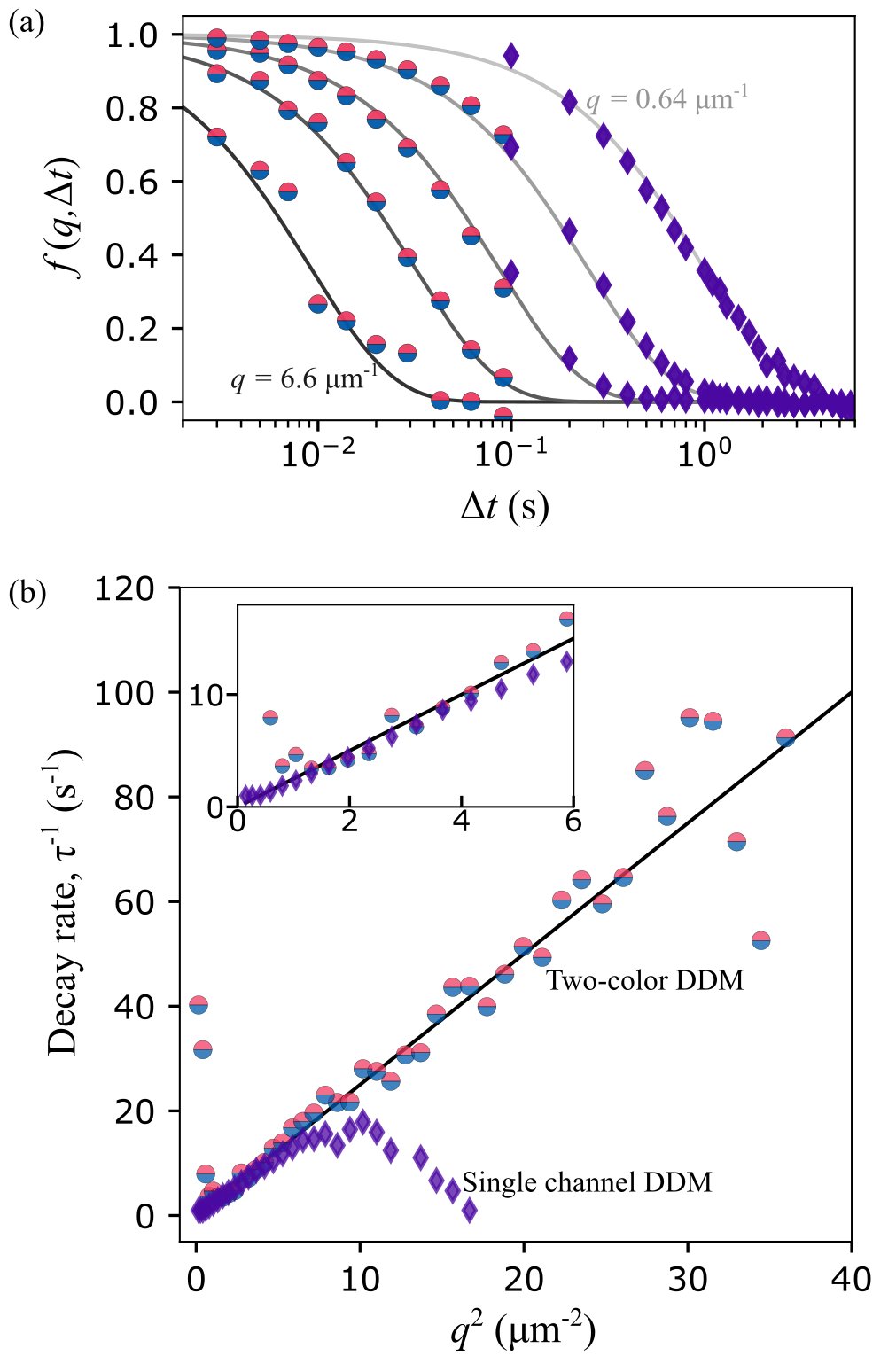}
\caption{\label{fig:fig2} (a) For five different values of the wavevector $q$ we plot  $f(q,\Delta t)$. In blue and red split circles, the values of $f(q,\Delta t)$ are shown for 10 lag times corresponding to the time intervals between the blue and red illumination pulses that occur within a single camera frame. These lag times range from 3 to 91 ms. In violet diamonds, the values of $f(q,\Delta t)$ are shown using standard DDM methods on just the blue channel of the acquired image sequence. Since the camera was running at 10 Hz, for standard DDM methods the minimum time lag corresponds to 100 ms. The solid lines represent the theoretical intermediate scattering function for diffusive dynamics, $f(q,\Delta t)=\exp(-Dq^2 \Delta t)$. (b) The determined decay rate, $\tau ^{-1}$, is plotted versus $q^2$. For diffusive dynamics, $\tau ^{-1}=Dq^2$ and the solid line here shows $D=2.48  \mu \mathrm{m}^2⁄\mathrm{s}$. Data from the standard DDM method (violet diamonds) starts deviating from this straight line just after the expected decay rate exceeds 10 s$^{-1}$, as anticipated given the images were acquired at 10 frames per second.  Data from the two-color DDM method follows the expected linear relationship up to around 70 s$^{-1}$. The inset focuses on the low-$q$ region. The two-color method does not accurately detect data fluctuations slower than about 10 s$^{-1}$.}
\end{figure}

After splitting this time series of images into blue and red channels, we find the two-color image structure function, $D_{two-color} (q,\Delta t)$. This two-color image structure function is found for the 10 time lags, $\Delta t$, that correspond to the delay times between the blue and red illumination pulses that occur within a single camera frame. With the same sequence of 8000 images we also find $D_{standard} (q,\Delta t)$ using only data from the blue channel of each frame. For this, the values of $\Delta t$ range from 100 ms (the time between frames) to 10 s.

For both $D_{two-color} (q,\Delta t)$ and $D_{standard} (q,\Delta t)$, we proceed with typical DDM analysis where for each $q$ we fit the image structure function to Equation 1 and find $A$, $B$ and $\tau$. In Fig. \ref{fig:fig2}a we show $f(q,\Delta t)$, and the relationship between $\tau$ and $q$ is shown in Fig. \ref{fig:fig2}b. We observe that our data follows the expected exponential decay of $f(q,\Delta t)$ whether using the two-color or standard method. However, for the two-color method, we are able to observe the dynamics at larger wavevectors. Our data is consistently fit for wavevectors with decay rates of up to about 70 s$^{-1}$. Such fast dynamics are inaccessible using standard DDM on data acquired at 10 Hz. In Fig. \ref{fig:fig2}b, we observe, as expected, that $\tau^{-1}$ depends linearly on $q^2$. Decay rates determined through $D_{standard} (q,\Delta t)$ start to deviate from this linear relationship beyond about 10 s$^{-1}$. However, with $D_{two-color} (q,\Delta t)$ we find decay rates that match the expected linear relationship up to several times faster than the rate at which images were acquired. 

The decay rates, $\tau^{-1}$, determined from the two-color DDM method appear noisier than those from the standard method. This is likely the result of less data going into the two-color DDM method. When fitting our data to Equation 1, for each value of $q$ we have ten data points (corresponding to the 10 time intervals between blue and red illumination pulses) going into $D_{two-color} (q,\Delta t)$ and 30 data points contributing to $D_{standard} (q,\Delta t)$ (corresponding to lag times logarithmically spaced from 100 ms to 10 s). Furthermore, for each lag time, $\Delta t$, we average together 800 Fourier transformed image differences for $D_{two-color} (q,\Delta t)$. Whereas with the standard DDM method applied to the same data set, we have 7999 Fourier transformed image differences to average for $\Delta t$ of one frame (100 ms). 

We note that modifications to this two-color method could be made. For example, an image splitter system could be used to separate the two colors onto distinct regions of the camera sensor which would allow for the use of monochromatic cameras. Furthermore, while we applied this method to bright-field imaging, one could use similar principles using fluorescence with spectrally distinct fluorophores. 

In summary, we have devised a method to allow differential dynamic microscopy to be used to study dynamics faster than the camera frame rate. This obviates the need for high speed cameras to measure fast dynamics if one can instead illuminate the sample with spectrally-separated pulses offset in time and then split the different color signals on the image sensor. We show that this two-color DDM method allows us to extract the dynamics of colloidal particles that occur up to several times faster than the camera frame rate. \newline

R.M. acknowledges support from the Research Corporation for Science Advancement through the Cottrell Scholars program. \newline

The data that support the findings of this study are available from the corresponding author upon reasonable request.

\nocite{*}
\bibliography{twocolorddm}

\end{document}